\documentclass{llncs} 

\author{Jean-Marc Rosengard \and  Marian F. Ursu}
\institute{University of London\\
\email{http://w2.syronex.com/jmr/}}
\title{Ontological Representations of Software Patterns}

\begin{document}
\maketitle

\begin{abstract}
This paper\footnote{Published in the proceedings of KES'04, Lecture Notes in Computer Science, vol. 3215, pp. 31-38, Springer-Verlag, 2004.} is based on and advocates the trend in software engineering of extending the use of software patterns as means of structuring solutions to software development problems (be they motivated by best practice or by company interests and policies). The paper argues that, on the one hand, this development requires tools for automatic organisation, retrieval and explanation of software patterns. On the other hand, that the existence of such tools itself will facilitate the further development and employment of patterns in the software development process. The paper analyses existing pattern representations and concludes that they are inadequate for the kind of automation intended here. Adopting a standpoint similar to that taken in the semantic web, the paper proposes that feasible solutions can be built on the basis of ontological representations.
\end{abstract}

\section{Introduction}

Software patterns are proven solutions to recurring software construction problems in a given context. They describe the knowledge refined from experienced practitioners about a particular aspect of the domain of discourse. The concept of design pattern was formulated in \cite{Gamm93b} and has since been generalised in for example \cite{cope96}. The practical employment of patterns in software development has continuously grown \cite{dp10yrs}.

Patterns are generally intended for human/manual use, as structured but informal documentations. Their primary aim is to guide software engineers, by presenting and explaining solutions regarding software construction applicable to a particular context. They are means of structuring solution descriptions. From this angle, therefore, they could be regarded as \emph{codes of good practice}. This is the perspective that we take in this paper. In this context, patterns are presented in printed catalogues, employing mainly natural language, but also examples of code and diagrammatic representations.

Patterns can occur at different \emph{levels of abstraction} with respect to the specification of a software solution. Thus, there may be architectural patterns, design patterns in a specific design paradigm (e.g. \emph{observer} and \emph{composite} \cite{GoF} in the OO paradigm) and language-specific patterns (e.g. \emph{counted body} \cite{cope92} in C++, and \emph{marker interface} in Java). The latter can also be called idioms. Furthermore, software patterns may have different degrees of generality. Some may be application or domain specific (e.g., \emph{contract} and \emph{portfolio} \cite{anapat} in trading applications), whereas others may be general design patterns, applicable across application domains (e.g. \emph{observer} and \emph{composite}).

In recent years, software development using patterns has become common practice in the practitioners' community \cite{dp10yrs}. Subsequently, the amount of refined patterns is growing, irrespective of their category---from general language-specific patterns to application specific patterns. However, it is towards the application-domain end that a high rate of growth is expected. The amount of printed documentation, thus, too, is increasing, to the extent that it becomes difficult for it to be effectively used. The problems that appear in this context are similar to the problems faced by engineering designers who have to comply with large codes of regulations and good design practice, which we have already discussed in \cite{ursu98}.

We aim to develop tools for \emph{intelligent dissemination} of patterns to software practitioners. We propose a general model that is applicable to patterns disregarding their level of abstraction (specification) and generality (application domain). On its basis we will implement specific solutions for different categories of patterns. We are here adapting some of the solutions we have proposed previously for the dissemination and enforcement of regulatory engineering-design knowledge (e.g., \cite{ursu98}) to the domain of software engineering and software patterns.

Research has been investigating the possibility of automatic code generation from formal representations of software patterns \cite{1997:ase:eden}. The goal, according to this approach, is to reduce as much as possible the involvement of the human agent from the design and implementation process. This may be feasible for restricted areas. However, our standpoint is to develop tools that \emph{empower} rather than replace the software practitioners; ``\emph{patterns should not, cannot and will not replace programmers}'' \cite{cope96}. This is consistent with our previous work in intelligent design \cite{ursu00}. 

\section{Software Patterns: Evolution}
\label{sec:evol}

In their early years, patterns have been mainly used within the community close to the group that described them. A number of fundamental patterns have been refined, in particular at the level of design \cite{GoF}, and are now widely used among software engineers. They are involved in the construction of most medium-size and large object-oriented systems. Some have also been integrated in programming platforms, such as Java, becoming thus readily available for application programming.

As the result of almost a decade of pattern mining, a \emph{large quantity} (hundreds) of patterns have been described, reviewed, and catalogued. However there have been few initiatives to structure and organise this knowledge \cite{dp10yrs} into a consistent representation framework.

The rate of growth varies with respect to the level of abstraction---with reference to the specification of a solution---but more so with the level of generality---with reference to the reusability across application domains. General or core patterns tend to be considered as fundamental abstractions and, in time, become integrated into programming languages and tools. Their number is limited and essentially constant. A rate of growth is displayed by patterns specified at the level of middleware. This is because software applications are increasingly complex and, thus, have to be developed around middleware platforms (e.g. J2EE).

A higher rate can be predicted at the level of particular application domains or within particular software development companies. Patterns can naturally describe expertise regarding a specific software development application. Furthermore, they can also naturally express specific policies regarding software development within different organisations. The  focus, here, is on promoting the use of explicit, locally defined constructs, motivated by concerns like quality, security, performance or code management.

Domain specific patterns is the category that strongly motivates our work. Because they represent a natural way for the formulation of accumulated expertise and policies, we consider that they will become the means for the representation of such knowledge. Consequently, large \emph{knowledge repositories} of domain specific patterns will be created (both within organisations and for general use). 
Furthermore, domain specific patterns form a dynamic pool of knowledge. They are expected to evolve more rapidly than the more generic ones, because the requirements within application domains are under continuous change, and their review and publication process can be expected to be less rigorous.

At this end, manual use of patterns is not an effective solution anymore. Their expected development depends on the existence and motivates the development of tools for their automatic organisation, retrieval and explanation. By development we mean both \emph{refinement/statement} and \emph{employment/use}. The latter term encapsulates all the various cognitive activities involved in software development---such as understanding whether to use or not a pattern applicable to a given context, choosing a particular pattern suitable to a specific context and understanding how to generate code in accordance to a particular chosen pattern---and sharing.

\section{Existing Pattern Representations}
This section discusses existing representations of patterns and their suitability to automatic organisation, retrieval and provision of explanations. 

\subsection{Informal Representation}
Patterns are most generally represented in natural language, and are typically published in printed catalogues. The term ``presentation'' seems more suitable for this type of description. Such documents are loosely structured, in what we call canonical forms. Such a structure consists of a series of fields, each having a meaning introduced via an informal definition or description. An example of a canonical form is that proposed in \cite{GoF}. A fragment of this is illustrated in Table \ref{tbl:informal}, below.

\begin{table}[ht]
\caption{\label{tbl:informal}Fragment of a canonical form for pattern representation \cite{GoF}.}
\begin{tabular}{ll}
\hline\noalign{\smallskip}
{\bf Field} & {\bf Explanation / Definition} \\
\noalign{\smallskip}
\hline
\noalign{\smallskip}
Name 
& Ideally a meaningful name that will be part of the shared design\\
& vocabulary. Many existing patterns do not satisfy this requirement\\
& for historical reasons.\\ 
Also known as 
& Other names of the pattern. \\ 
Intent 
& A short specification or rationale of the pattern, used as a principal\\
& index for goal-oriented pattern search.\\ 
Applicability 
& An outline of the circumstances in which the pattern may be appli-\\
& cable and, perhaps more importantly, when it should not be applied.\\ 
Structure 
& A diagrammatic representation of the pattern.\\ 
Consequences 
& Discusses the context resulting from applying the pattern. In parti-\\
& cular, trade-offs should be mentioned.\\ 
Implementation 
& Advices on how to implement the patterns, and other language spe-\\
& cific issues. The implementation will depend on the abstractions \\
& (objects, parameterised types,\ldots) supported by the target language. \\ 
Known uses 
& Patterns are by essence derived from existing systems. It is therefore \\
& important that they be justified by their use in several real systems. \\ 
Related patterns 
& Patterns are often coupled or composed with other patterns, leading\\
& to the concept of pattern language; e.g. a visitor may be used to\\
& apply an operation to the closed structure provided by a composite.\\ 
\hline
\end{tabular}
\end{table}

Consider, for example, the most common situation when a software developer is within a specific situation and wants to identify whether there exists a particular pattern useful to the situation at hand. A search within a repository of patterns would, most probably, involve the \emph{intent} and \emph{applicability} descriptors. Assuming that the catalogue exists in an electronic format that preserves the structure of the printed catalogue, as described above, then the best option available to him is a keyword search; intent and applicability have no internal structures. This means that the software engineer attempts to retrieve documents describing relevant patterns based on phrases that he would have to guess are used in the descriptors. Each time a document/pattern is retrieved, he would have to read it thoroughly---since no summarisation or explanatory features would be supported by the discussed representation---and decide upon its suitability. Obviously, this is a cumbersome process.

The drawbacks of such a retrieval process are well known. They are more critical if the agent who carries out the process does not have at least some knowledge of the descriptions' jargon or of the possible expected results; in our case, if the software engineer is a novice. Note that by {\em novice}, we mean without much software development experience, but also inexperienced with a particular technology, or new to a company and not familiar with its policies and codes. 

 These drawbacks have been identified and well described in the context of the web and represent a motivating factor for the development of the \emph{semantic web} \cite{hiit02}. 

Although they have the same nature, the scale of the problem in the context of software patterns is obviously not as large as in the context of the web. However, the effects can be similarly drastic, under the founded assumption that the pattern repository has a substancial size. Furthermore, missing or misusing a pattern with respect to a particular situation could have severe implications if the patterns represent company policies or codes of best practice. 
The above argument was implicitly carried out at the level of application/domain specific patterns. However, it is equally valid in the context of domain independent, but language-specific patterns (idioms). A good example for this is the Javadoc documentation of the Java platform. This knowledge base is significantly large and finding relevant solutions to a specific problem is a cumbersome process for non-expert Java programmers.
 
Another major drawback of this representation is the fact that it does not readily support knowledge management and sharing, also necessarily required, in particular for application-domain patterns (refer to Section \ref{sec:evol}). Informal representations based on canonical forms cannot support the level of automation at which we aim. For this, we need better-structured representations.

\subsection{Patterns in UML}

Patterns are represented in UML using the familiar class/object and interaction diagrams, and also using the more specific \emph{parameterised collaboration} model \cite{umlref}---allowing the variation of roles around a given collaboration. While these representations are useful for understanding a pattern and guiding through its implementation, they only express the \emph{structural} aspects of the pattern. They do little to help the engineer understand its higher-level concerns, like its intent, applicability and tradeoffs. Unsurprisingly, UML is not suitable for pattern representation for the purpose stated in this paper. As a simple example, consider the \emph{strategy} and \emph{state} patterns. Although their intents \cite{GoF} are very different, they exhibit a similar structure. 

\subsection{Formal Representations}

Although patterns primarily constitute a body of knowledge for human consumption, several initiatives have been made to formalise some aspects of their representation, opening the way to some automated support for pattern-based development.

Formalisation is applied to some of the essential properties of patterns (pattern invariants) by means of specification languages, like the Object Constraint Language (OCL) \cite{OCL97}. On the instantiation of a pattern or the modification of an existing occurrence of a pattern, its implementation may be automatically validated using the structural and behavioural/temporal constraints specified in OCL expressions. Such representations, although useful is such situations, cannot express all the knowledge (that a pattern encapsulates) required for instantiations or modifications. Furthermore, aspects like pattern intent, motivation and applicability, cannot be expressed in OCL. 

Constraint languages and ontologies (proposed here) are complementary in pattern representation. The former are focussed on automatic code generation, whereas the latter are focussed on the provision of \emph{intelligent advice} to software developers. Also, constraints and ontological representations operate at different stages---expressions represented in constraint languages become applicable after the engineer has made solution decisions. 

\section{Ontological Representation}

There are various meanings that the term \emph{ontology} can have in AI \cite{Uschold95}. We adopt the same view as \cite{gruber93} and take ontology first to mean a specification of a conceptualisation, and second---pragmatically---to define a (standard) vocabulary with which queries and assertions are exchanged among agents. Ontologies are particularly useful in knowledge sharing and reuse. If two agents make the same ontological commitment, then they can exchange knowledge. Alternatively, any knowledge base represented within an ontology can be accessed by agents who committed to the respective ontology. The latter viewpoint is relevant to our proposal. If an ontology for the representation of patterns is in place, then pattern repositories (represented in the respective ontology) become accessible by various tools---for intelligent organisation, retrieval and provision of explanations---provided they committed to the ontology. 

An ontology becomes standard within a community when sufficient (or sufficiently powerful) agencies from that community commit to it. The quality of an ontology being standard is only required if knowledge reuse and sharing is an issue within the community. \emph{sharing} and \emph{reuse} should be understood, in the context of software patterns, with respect to the type of the patterns. Idioms should be shareable across application domains, whereas application specific patterns may need to be shared only at the level of an institution.

As a method of work, we started with the development of a basic ontology for design patterns. They are of a manageable size and their generality is implicitly transferred to the basic ontology. Thereafter, we shall enhance the basic ontology with language-specific concepts and domain-specific concepts, when we move towards the representation of the respective software patterns. 

Although we do not necessarily intend that the deployment of documentation based on patterns be made within the web, our work, here, is strongly connected with that carried out within the \emph{semantic web} \cite{Fensel02}. 
The use of ontologies was proposed in software engineering, but in the context of component based development. The focus of these efforts (e.g., \cite{MIT:oxy}) is on automatic retrieval and assembly. Our focus is on the provision of intelligent advice to software engineers.  

\section{Conclusions}

In this paper, we introduced the idea of combining software patterns with ontological representations, with a view to developing tools for the automatic organisation, retrieval and explanation of reusable solutions to software development, codes of good practice and company policies.

\end{document}